\documentclass[american,aps,pra,reprint, superscriptaddress]{revtex4-1}
\usepackage[unicode=true,pdfusetitle, bookmarks=true,bookmarksnumbered=false,bookmarksopen=false, breaklinks=false,pdfborder={0 0 0},backref=false,colorlinks=false] {hyperref}
\hypersetup{ colorlinks,linkcolor=myurlcolor,citecolor=myurlcolor,urlcolor=myurlcolor}
\usepackage{graphics,epstopdf,graphicx, amsthm, amsmath, amssymb, times, braket, colortbl, color, bm, framed, cleveref, mathrsfs}
\usepackage[up]{subfigure}
\usepackage{tikz}
\definecolor{myurlcolor}{rgb}{0,0,0.7}

\newcommand{\setft}[1]{\mathrm{#1}}
\newcommand{\density}[1]{\setft{D}\left(#1\right)}

\newcommand{\proj}[1]{| #1\rangle\!\langle #1 |}

\newcommand{\Br}[1]{\left[#1\right]}

\theoremstyle{plain}
\newtheorem{thm}{Theorem}

\newtheorem{prop}[thm]{Proposition}

\newcommand*{\myproofname}{Proof}

\def\ot{\otimes}

\def\cH{\mathcal{H}}

\makeatother

\begin{document}

 \author{Kaifeng Bu}
 \email{bkf@zju.edn.cn}
 \affiliation{School of Mathematical Sciences, Zhejiang University, Hangzhou 310027, PR~China}
 \author{Asutosh Kumar}
 \email{asukumar@hri.res.in}
 \affiliation{Harish-Chandra Research Institute, Chhatnag Road, Jhunsi, Allahabad 211019, India}
 \author{Junde Wu}
 \email{wjd@zju.edn.cn}
 \affiliation{School of Mathematical Sciences, Zhejiang University, Hangzhou 310027, PR~China}

\title{Bell-type inequality in quantum coherence theory as an entanglement witness}

\begin{abstract}
Bell inequality is a mathematical inequality derived using the assumptions of locality and realism. Its violation guarantees the existence of quantum correlations in a quantum state. Bell inequality acts as an entanglement witness in the sense that a pure bipartite quantum state, having nonvanishing entanglement, always violates a Bell inequality.
We construct Bell-type inequalities for product states in quantum coherence theory for different measures of coherence, and find that the maximally entangled states violate these inequalities. We further show that Bell-type inequalities for relative entropy of coherence is violated by all two-qubit pure entangled states, serving as an entanglement witness.
\end{abstract}

\maketitle

\section{Introduction}
Quantum information theory is the issue of marriage of quantum mechanics and information theory. Quantum information protocols, due to presence of exotic resources like entanglement, can perform certain information-processing and communication tasks far efficiently than classical information protocols \cite{Nielsen00}.
Quantum entanglement \cite{HorodeckiRMP09} is a basic resource for various information processing protocols, such as
superdense coding \cite{Bennett1992} and teleportation \cite{Bennett1993}.
It is a useful resource due to its vast applicability in quantum computation and communication tasks \cite{Note1}, as well as in other information processing protocols \cite{Antonio2007,Pironio2009,Gisin2010,Masanes2011,Lo2012,Zukowski1998,Hillery1999,Demkowicz2009,Cleve1999,Karlsson1999,Calderbank1996,Steane1996,Steane1996,Calderbank1997}.
As put by Schrodinger, \emph{entanglement is not one but the characteristic trait of quantum mechanics} \cite{Schrodinger1935}. A composite state of a two-party system is said to be entangled if it cannot be written as a tensor product of states of the individual subsystems. 
The most important characteristic of quantum entanglement is its nonlocality \cite{Bell1964}, which has intrigued a lot of research in this area \cite{Buhrman2010,Clauser1978,Home1991,Khalfin1992,Brunner2014}.
One method to study the nonlocality is to build Bell inequality, which was first proposed by Bell in Ref.  \cite{Bell1964}, and later developed in Refs. \cite{Clauser1969,Fine1982,Gisin1991}.
Bell had constructed a mathematical inequality derived using the assumptions of locality and realism, and showed that it is violated by certain entangled quantum mechanical states \cite{Bell1964,Clauser1969}.
Violation of one or more Bell inequalities \cite{Brunner2014} ensure the existence of quantum correlations in a quantum
state. Moreover, they form necessary and sufficient criteria to detect entanglement in pure bipartite states \cite{Gisin1991,Gisin92}. A pure bipartite quantum state, having nonvanishing quantum correlation, always violates a Bell inequality. On the other hand, it is not necessarily true for bipartite mixed states and multipartite systems. For instance, Werner states \cite{Werner89}, for certain parameter ranges, do not violate the Clauser-
Horne-Shimony-Holt (CHSH) Bell inequality \cite{Clauser1969}. Similarly, there are examples of multiparty pure entangled states which do not violate multipartite correlation function \cite{Weinfurter2001,Werner2001,Zukowski2002}  Bell inequalities with binary measurement settings at each site \cite{Zukowski2002b,Sen2002}.

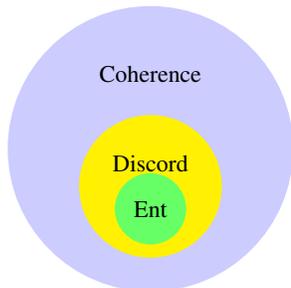
\begin{figure}[ht]
  \centerline{
    \begin{tikzpicture}[thick]
    \tikzstyle{operator} = [draw,fill=white,minimum size=1.5em]
   \tikzstyle{phase1} = [fill=blue!20,shape=circle,minimum size=12em,inner sep=0pt]
    \tikzstyle{phase2} = [fill=yellow,shape=circle,minimum size=6em,inner sep=0pt]
    \tikzstyle{phase3} = [fill=green!60,shape=circle,minimum size=3em,inner sep=0pt]
    \tikzstyle{surround} = [fill=blue!10,thick,draw=black,rounded corners=2mm]
    \tikzstyle{block} = [rectangle, draw, fill=white,
    text width=3em, text centered, , minimum height=6em]
    \node[phase1] (p1) at (5,3) {~};
    \node[phase2] (p2) at (5,2.5) {~};
    \node[phase3] (p3) at (5,2.2) {~};
    \node at (5,4) (e1){Coherence};
    \node at (5,2.8) (e2){Discord};
  \node at (5,2.2) (e3){\small {Ent}};
    \end{tikzpicture}
  }
  \caption{The relationship between entanglement, discord and coherence in composite quantum states \cite{Yao2015a}, where ``Ent'' denotes entanglement. The hierarchical relation, ${\cal C}(\rho) \geq {\cal D}(\rho) \geq {\cal E}(\rho)$, signifies that quantum coherence is a rather ubiquitous manifestation of quantum correlations.}
  \label{fig-re1}
\end{figure}

Like entanglement, several other useful quantum resources like
purity \cite{Horodecki2003}, reference frames \cite{Gour2008,Marvian2013}, thermodynamics \cite{Brandao2013,Gour2013}, etc. have been identified and
quantified thus far. The development of any resource theory requires identifying two basic components
\cite{Devetak2008,HorodeckiIJTPB2013,Coecke2016,Brandao2015b}, namely free states and free (or allowed) operations. Other states and operations are dubbed as a resource in the corresponding resource theory. For entanglement theory, separable states are free states and local operations and classical communication (LOCC) are free operations as these operations do not create entanglement on an average.
Recent progress  in the fields like quantum biology
\cite{Plenio2008,Levi14,Aspuru2009, Lloyd2011, Huelga13} and quantum thermodynamics \cite{Horodecki2013, Skrzypczyk2014, Narasimhachar2015, Brandao2015, Rudolph214, Rudolph114, Gardas2015, Avijit2015} reveal
the particular role of ``quantum coherence'' in quantum processing.
Coherence is identified by the presence of off-diagonal terms in the density matrix, and is a basis-dependent quantity. That is, a quantum state may have nonvanishing coherence in one basis, and vanishing coherence in another basis. 
Unlike entanglement and other quantum correlations such as quantum discord \cite{Henderson2001,Ollivier2001}, coherence is defined for both single and multipartite systems.
Baumgratz et al. in Ref. \cite{Baumgratz2014} have provided a quantitative theory of coherence as a new quantum resource.
To quantify coherence in a given state, several kinds of coherence measures such as $l_1$-norm of coherence and relative entropy of coherence have been proposed in \cite{Baumgratz2014,Girolami14}.
Quantum coherence, being a resource in quantum information and quantum thermodynamics, has arrested a lot of attention of the quantum community.
Besides quantification of quantum coherence, relationship between coherence and other
quantities like quantum entanglement, quantum discord,
mixedness are also established in \cite{Alex15,Uttam2015,Asutosh15,Xi2015,Yao2015a,Cheng2015}.
In Ref. \cite{Yao2015a}, authors investigated quantum coherence in multipartite systems and revealed the connection between coherence, quantum discord, and entanglement (see Fig~\ref{fig-re1}). Moreover, they also shown that the global unitary operations do not outdo than the local unitary operations.
That is, for the state $\ket{00}$ in $\mathcal{H}_A\ot \mathcal{H}_B$ with $dim \mathcal{H}_A=d_A$ and $dim \mathcal{H}_B=d_B$, 
it is easy to show that for any global unitary operation $U$,
there always exists a local unitary operation $V_1\ot V_2$ such that
\begin{eqnarray}
\mathcal{C}_{l_1}(U\ket{00})=\mathcal{C}_{l_1}(V_1\ot V_2\ket{00}),
\end{eqnarray}
which is very different from the entanglement case as local unitaries can not increase the entanglement. Thus, it seems that the global unitary operations can not be better than local unitary operations for quantum coherence.
Other developments like the freezing phenomena \cite{Bromley2015}, the coherence transformations under incoherent operations \cite{Du2015}, establishment of geometric lower bound for a coherence measure \cite{Pires2015}, creation of coherence
using unitary operations \cite{Yao2015a}, energy cost of creating quantum coherence \cite{Avijit2015} have also been reported.
On the other hand, authors in Ref. \cite{Winter2015} have proposed an operational resource theory of quantum coherence, which provides physical interpretations to the coherence measures via the transformation processes like coherence distillation.

In this paper, we investigate the quantum coherence in multipartite systems, and pursue the answers to the following questions:
(i) Can we construct Bell-type inequalities in quantum coherence theory to demonstrate the nonlocal properties present in quantum coherence? (ii) Do the violations of these inequalities guarantee the existence of quantum correlations in a quantum state? We find that the answer is in affirmative. We construct Bell-type inequalities for product states for both $l_1$-norm of coherence and relative entropy of coherence based on the idea of Bell inequality in entanglement theory. We find that the maximally entangled state violates these inequalities. We further show that the Bell-type inequality for relative entropy of coherence is violated by all two-qubit pure entangled states, thus acting as an entanglement witness.

The paper is organized as follows. We start with providing preliminaries on measures of coherence and Bell inequality in Sec. \ref{sec:prelims}.
In Sec. \ref{sec:bell-ineq}, we construct Bell-type inequalities for bipartite product states for different measures of coherence, and give examples to show that these inequalities are violated by the maximally entangled two-qubit pure state. We further show that the Bell-type inequality for relative entropy of coherence is violated by all two-qubit pure enatngled states, serving as an entanglement witness in Sec. \ref{sec:ent-witness}. Finally, we conclude in Sec. \ref{sec:con} with a discussion on the results obtained.

\section{Preliminaries}
\label{sec:prelims}
In this section, we briefly recall the valid measures of coherence that quantify quantum coherence in quantum states in a fixed reference basis, and discuss about a particular Bell inequality called the Clauser-Horne-Shimony-Holt (CHSH) inequality.

\subsection{Measures of Coherence}
For a quantum state $\rho$, in the reference basis
$\{|i\rangle\}$, two proper measures of quantum coherence in the framework of Ref. \cite{Baumgratz2014}, are $l_1$-norm of coherence ($\mathcal{C}_{l_{1}}$) and relative entropy of coherence ($\mathcal{C}_{r}$). They are defined as following:
(i) The measure of coherence, $\mathcal{C}_{l_{1}}$, based on $l_1$-norm is defined by
\begin{eqnarray}
\mathcal{C}_{l_{1}}(\rho):=\sum_{i\neq j}|\rho_{ij}|.
\end{eqnarray}
For the pure state, $|\Psi\rangle = \sum_i c_i |i\rangle$, it reduces to
\begin{eqnarray}
\mathcal{C}_{l_{1}}(|\Psi\rangle):=\big(\sum_{i}|c_i|\big)^2 - 1,
\end{eqnarray}
where $c_i = \langle i|\Psi\rangle$.
(ii) The measure of coherence, $\mathcal{C}_{r}$, based on relative entropy is defined by
\begin{eqnarray}
\mathcal{C}_{r}(\rho):= \mathrm{min}_{\sigma \in {\cal I}} S(\rho \parallel \sigma) = S(\rho^d) - S(\rho),
\end{eqnarray}
where ${\cal I}$ is the set of all incoherent states in the reference basis $\{|i\rangle\}$,
$S(\rho \parallel \sigma) = \mathrm{Tr}{\rho(\log\rho - \log\sigma)}$ is the relative entropy between $\rho$ and $\sigma$, $S(\rho)=-\mathrm{Tr}{\rho\log\rho}$ is the von Neumann entropy of $\rho$, and $\rho^d$ is the diagonal state of $\rho$, i.e., $\rho^d = \sum_i \langle i|\rho|i\rangle |i\rangle \langle i|$.
For the maximally coherent pure state given by $|\Psi_d\rangle = \frac{1}{\sqrt{d}}\sum_{i=0}^{d-1}|i\rangle$ \cite{Baumgratz2014}, $\mathcal{C}_{l_{1}}(|\Psi_d\rangle) = d-1$ and $\mathcal{C}_{r}(|\Psi_d\rangle) = \log~d$.

\subsection{Bell Inequality}
Bell, in 1964, constructed an inequality (referring to correlations of measurement results) that is satisfied
by all theories that are both local and counterfactual-definite \cite{Bell1964}. He then showed that quantum
mechanics violates this inequality. That is, quantum mechanics cannot be both local and counterfactual-
definite. This is the celebrated Bell's theorem.
Thus, based on the assumptions of locality and reality, one can construct mathematical relations, the CHSH-Bell inequalities \cite{Bell1964,Clauser1969}, which can be violated by quantum mechanical systems.
Consider two observers Alice and Bob share a bipartite state $\rho_{AB}$. Suppose that Alice has two observables \(Q\) and \(R\) with two possible outcomes \(\pm 1\) for each observable. Similarly, Bob has two observables \(S\) and \(T\) with two possible outcomes \(\pm 1\) again for each observable. In the classical case (that is, if a local hidden variable model exists for this state), it is easy to show that for these four dichotomic observables, we have
\begin{eqnarray}
|E(QS+RS+RT-QT)|\leq 2.
\end{eqnarray}
However, this inequality can be shown to be violated by several quantum mechanical two-qubit bipartite states. In quantum mechanics, we have
\begin{eqnarray}
|E(QS+RS+RT-QT)|\leq 2\sqrt{2}.
\end{eqnarray}
The maximum value is achieved when Alice and Bob share the EPR pair, $|\Psi\rangle = \frac{1}{\sqrt{2}}(|01\rangle - |10\rangle)$, and their observables are
\begin{eqnarray*}
Q=Z,&&~~R=X\\
S=(-Z-X)/\sqrt{2},&&~~T=(Z-X)/\sqrt{2}
\end{eqnarray*}
where
$ X= \Br{\begin{array}{ccc}
0 & 1  \\
1 & 0
\end{array}}
$
and
$ Z= \Br{\begin{array}{ccc}
1 & 0 \\
0 & -1
\end{array}}
$
are Pauli matrices. For more details on Bell inequality and nonlocality, see Ref. \cite{Brunner2014}.

\section{Bell-type inequality for Product states in Quantum Coherence}
\label{sec:bell-ineq}
In this section, we consider coherence in bipartite quantum systems.
In particular, we build Bell-type inequalities for bipartite product states for different measures of coherence, and find that two-qubit maximally entangled pure states violate these inequalities. We further show that Bell-type inequality for relative entropy of coherence is violated by all two-qubit entangled pure states, acting as an entanglement witness.
We are also able to find a basis in which Bell-type inequality for $l_1$-norm of coherence is violated by a large number of two-qubit pure entangled states.
Since coherence is a basis-dependent quantity, we must specify the basis we have chosen.
As we know, \emph{if we choose observable X in $\mathcal{H}_A$ and observable Y in $ \mathcal{H}_B$, then the eigenstates of X and Y can form a local basis of $\mathcal{H}_A\ot \mathcal{H}_B$}.
For a bipartite state $\rho_{AB}$, and any two observables X and Y in $\mathcal{H}_A$ and $\mathcal{H}_B$ respectively, we will denote $\mathcal{C}_r(X,Y,\rho_{AB})$ and $\mathcal{C}_{l_1}(X,Y,\rho_{AB})$ by $\mathcal{C}_r(\rho_{AB})$ and $\mathcal{C}_{l_1}(\rho_{AB})$ in the local basis formed by X and Y. Likewise, $\mathcal{C}_r(X,\rho_{A})$ and $\mathcal{C}_{l_1}(X,\rho_{A})$ will be denoted by $\mathcal{C}_r(\rho_A)$ and $\mathcal{C}_{l_1}(\rho_A)$ respectively in the local basis formed by X. $\mathcal{C}_r(Y,\rho_{B})$ and $\mathcal{C}_{l_1}(Y,\rho_{B})$ are denoted similarly.

\subsection{Bell-type inequality for $l_1$-norm of coherence}
\label{sec:l1case}

Here we prove that there exists a Bell-type inequality for $l_1$-norm of coherence for bipartite product states. The violation of this inequality would then suggest that quantum states are entangled, and hence nonlocal.

\begin{thm}
In two-qubit system $\cH_A\ot\cH_B$, if the global state $\rho_{AB}$ is a product state $\rho_A\ot\rho_B \in \density{\cH_A\ot\cH_B}$, then for any observables R, Q in $\cH_A$ and observables S, T in $\cH_B$, we have
\begin{eqnarray}
&&[\mathcal{C}_{l_1}(Q,S,\rho_{AB})+1]+[\mathcal{C}_{l_1}(R,S,\rho_{AB})+1]\nonumber\\
&+&[\mathcal{C}_{l_1}(R,T,\rho_{AB})+1]
-[\mathcal{C}_{l_1}(Q,T,\rho_{AB})+1]\leq 8.\nonumber
\end{eqnarray}
\end{thm}

\begin{proof}
When $\rho_{AB}=\rho_A\ot\rho_B$, we have the following fact
 \begin{eqnarray}
 \label{eq:cl1-norm}
 \mathcal{C}_{l_1}(X,Y,\rho_A\ot\rho_B)+1=(\mathcal{C}_{l_1}(X,\rho_A)+1)(\mathcal{C}_{l_1}(Y,\rho_B)+1).
 \nonumber\\
 \end{eqnarray}
Note that this is true for any product state in arbitrary dimensions.

Then, using (\ref{eq:cl1-norm}), we have
\begin{eqnarray*}
&&[\mathcal{C}_{l_1}(Q,S,\rho_A\ot\rho_B)+1]+[\mathcal{C}_{l_1}(R,S,\rho_A\ot\rho_B)+1]\\
&+&[\mathcal{C}_{l_1}(R,T,\rho_A\ot\rho_B)+1]
-[\mathcal{C}_{l_1}(Q,T,\rho_A\ot\rho_B)+1]\\
&&=(\mathcal{C}_{l_1}(Q,\rho_A)+1)(\mathcal{C}_{l_1}(S,\rho_B)+1)\\
&&+(\mathcal{C}_{l_1}(R,\rho_A)+1)(\mathcal{C}_{l_1}(S,\rho_B)+1)\\
&&+(\mathcal{C}_{l_1}(R,\rho_A)+1)(\mathcal{C}_{l_1}(T,\rho_B)+1)\\
&&-(\mathcal{C}_{l_1}(Q,\rho_A)+1)(\mathcal{C}_{l_1}(T,\rho_B)+1)
\end{eqnarray*}
For simplicity, let us write
\begin{eqnarray*}
W_Q=\mathcal{C}_{l_1}(Q,\rho_A)+1,~~W_R=\mathcal{C}_{l_1}(R,\rho_A)+1\\
W_S=\mathcal{C}_{l_1}(S,\rho_B)+1,~~W_T=\mathcal{C}_{l_1}(T,\rho_B)+1
\end{eqnarray*}
For two-dimensional systems, we have $1\leq W_X\leq 2$, for $X=Q,~R,~S,~T$.
Hence, the absolute value of above equation is given by
\begin{eqnarray*}
&&|W_Q W_S+W_R W_S+W_R W_T-W_Q W_T|\\
&=&|W_Q(W_S-W_T)+W_R(W_S+W_T)|\\
&\leq& W_Q|W_S-W_T|+W_R|W_S+W_T|\\
&\leq& 8.
\end{eqnarray*}
That is,
\begin{eqnarray}
\label{eq:bell-cl1}
&&[\mathcal{C}_{l_1}(Q,S,\rho_A\ot\rho_B)+1]+[\mathcal{C}_{l_1}(R,S,\rho_A\ot\rho_B)+1]\nonumber\\
&+&[\mathcal{C}_{l_1}(R,T,\rho_A\ot\rho_B)+1]
-[\mathcal{C}_{l_1}(Q,T,\rho_A\ot\rho_B)+1]\leq 8.\nonumber\\
\end{eqnarray}
Hence the proof.
\end{proof}
The equality will be achieved when $W_Q=W_R=W_S=W_T=2$, i.e., when $\rho_A$ and $\rho_B$
are maximally coherent pure states \cite{Peng2015}.
\emph{We will denote the left-hand side of inequality (\ref{eq:bell-cl1}) by ${\cal B}_{{\cal C}_{l_1}}$}.

\subsection{Bell-type inequality for relative entropy of coherence}
\label{sec:rel}

In full analogy to the $l_1$-norm of coherence, we can also establish
Bell-type inequality for relative entropy of coherence.

\begin{thm}
In two-qubit system $\cH_A\ot\cH_B$, if the global state $\rho_{AB}$ is a product state $\rho_A\ot\rho_B \in \density{\cH_A\ot\cH_B}$, then for any observables R, Q in $\cH_A$ and observables S, T in $\cH_B$, we have
\begin{eqnarray}
&&\mathcal{C}_r(Q,S,\rho_{AB})+\mathcal{C}_r(R,S,\rho_{AB})\nonumber\\
&+&\mathcal{C}_r(R,T,\rho_{AB})-\mathcal{C}_r(Q,T,\rho_{AB})\leq 4.\nonumber
\end{eqnarray}
\end{thm}

\begin{proof}
When $\rho_{AB}=\rho_A\ot\rho_B$, it is easy to see that
\begin{eqnarray}
\label{eq:rel-ent}
\mathcal{C}_r(X,Y,\rho_A\ot\rho_B)=\mathcal{C}_r(X,\rho_A)+\mathcal{C}_r(Y,\rho_B).
\end{eqnarray}
This is also true for any product state in arbitrary dimensions.
Therefore, using (\ref{eq:rel-ent}), we have
\begin{eqnarray}
\label{eq:bell-cr}
&&\mathcal{C}_r(Q,S,\rho_A\ot\rho_B)+\mathcal{C}_r(R,S,\rho_A\ot\rho_B)\nonumber\\
&+&\mathcal{C}_r(R,T,\rho_A\ot\rho_B)-\mathcal{C}_r(Q,T,\rho_A\ot\rho_B)\nonumber\\
&=&2[\mathcal{C}_r(R,\rho_A)+\mathcal{C}_r(S,\rho_B)]\leq 4.
\end{eqnarray}
Hence proved.
\end{proof}
Similar to the $l_1$-norm case, the equality will be obtained when $\rho_A$ and $\rho_B$ are maximally coherent pure states. \emph{We will denote the left-hand side of inequality (\ref{eq:bell-cr}) by ${\cal B}_{{\cal C}_{r}}$}.

\subsection{Violation of Bell-type inequalities}
Here we show explicitly that Bell-type inequalities for both $l_1$-norm of coherence and relative entropy of coherence are violated by the maximally entangled (singlet) state, $|\Psi\rangle = \frac{1}{\sqrt{2}}(|01\rangle - |10\rangle)$, for some reference bases.\\

\noindent
\emph{Example 1.} Consider the basis
\begin{eqnarray*}
&&Q: \set{\ket{0}, \ket{1}},~~R: \set{\ket{R_+}, \ket{R_-}}\\
&&S: \set{\ket{+}, \ket{-}},~~T: \set{\ket{0}, \ket{1}},
\end{eqnarray*}
where $\ket{R_+}=\frac{1}{\alpha}(\ket{+}+\ket{0})$ and $\ket{R_-}=\frac{1}{\beta}(\ket{+}-\ket{0})$ are the eigenstates of the observable $(-Z-X)/\sqrt{2}$, and $\alpha=\sqrt{2+\sqrt{2}}$, $\beta=\sqrt{2-\sqrt{2}}$.
Then, after some simple calculations, we obtain $\mathcal{C}_{l_1}(Q,S,\ket{\Psi})+1=4$, $\mathcal{C}_{l_1}(R,S,\ket{\Psi})+1=\mathcal{C}_{l_1}(R,T,\ket{\Psi})+1=2+\sqrt{2}$ and $\mathcal{C}_{l_1}(Q,T,\ket{\Psi})+1=2$.
Hence,
\begin{eqnarray*}
{\cal B}_{{\cal C}_{l_1}}=6+2\sqrt{2} > 8.
\end{eqnarray*}
In fact, the above inequality is true for any two-qubit maximally entangled state.
Similarly, we obtain $\mathcal{C}_r(Q,S,\ket{\Psi})=2$, $\mathcal{C}_r(R,S,\ket{\Psi})=\mathcal{C}_r(R,T,\ket{\Psi})=H(\frac{1}{4\alpha^2}, \frac{1}{4\alpha^2}, \frac{1}{4\beta^2}, \frac{1}{4\beta^2})=1.6009$ and
$\mathcal{C}_r(Q,T,\ket{\Psi})=1$.
Therefore,
\begin{eqnarray*}
{\cal B}_{{\cal C}_{r}}=4.2018 > 4.
\end{eqnarray*}
Again, the above inequality is true for any two-qubit maximally entangled state.\\

\noindent
\emph{Example 2.} Compared to \emph{Example 1} above, we obtain larger violations of both Bell-type inequalities if we take the following basis

\begin{eqnarray*}
&&Q:\set{\ket{0}, \ket{1}},~~R:\set{\ket{R_+}, \ket{R_-}}\\
&&S: \set{\ket{S_+}, \ket{S_-}},~~T: \set{\ket{0}, \ket{1}},
\end{eqnarray*}
where
\begin{equation*}
\Br{\begin{array}{ccc}
\ket{R_+}&\ket{R_-}
\end{array}}
=\Br{\begin{array}{ccc}
\ket{0}&
\ket{1}
\end{array}}\Br{\begin{array}{ccc}
\frac{1}{\sqrt{2}}&\frac{i}{\sqrt{2}}\\
\frac{i}{\sqrt{2}}&\frac{1}{\sqrt{2}}
\end{array}},
\end{equation*}
and
\begin{equation*}
\Br{\begin{array}{ccc}
\ket{S_+}&
\ket{S_-}
\end{array}}
=\Br{\begin{array}{ccc}
\ket{0}&
\ket{1}
\end{array}}\Br{\begin{array}{ccc}
\frac{1}{\sqrt{2}}&\frac{1}{\sqrt{2}}\\
-\frac{1}{\sqrt{2}}&\frac{1}{\sqrt{2}}
\end{array}}.
\end{equation*}

We, then, have
$\mathcal{C}_{l_1}(Q,S,\ket{\Psi})+1=\mathcal{C}_{l_1}(R,S,\ket{\Psi})+1=\mathcal{C}_{l_1}(R,T,\ket{\Psi})+1=4,~~\mathcal{C}_{l_1}(Q,T,\ket{\Psi})+1=2$, and
$\mathcal{C}_r(Q,S,\ket{\Psi})=\mathcal{C}_r(R,S,\ket{\Psi})=\mathcal{C}_r(R,T,\ket{\Psi})=2,~~
\mathcal{C}_r(Q,T,\ket{\Psi})=1$.
Therefore,
\begin{eqnarray*}
{\cal B}_{{\cal C}_{l_1}}=10,
\end{eqnarray*}
and
\begin{eqnarray*}
{\cal B}_{{\cal C}_{r}}=5.
\end{eqnarray*}
\begin{figure}
\begin{center}
\includegraphics[width=2.4in, angle=0]{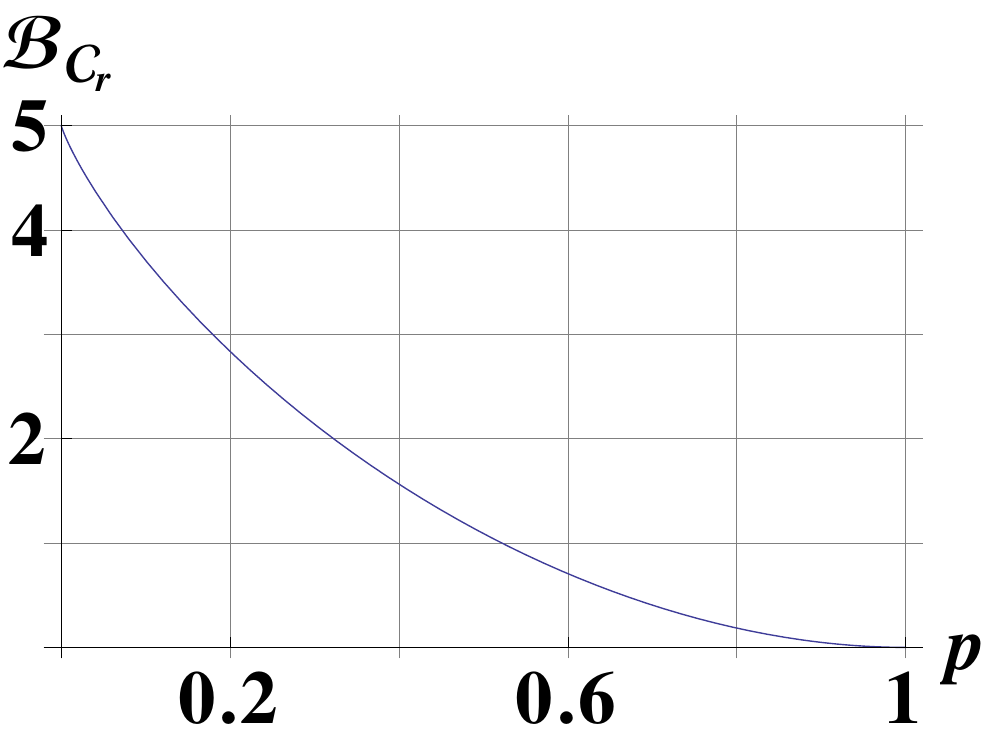}
\caption{Plot of the quantity, ${\cal B}_{{\cal C}_{r}}=\mathcal{C}_r(Q,S,\rho_{AB})+\mathcal{C}_r(R,S,\rho_{AB})+\mathcal{C}_r(R,T,\rho_{AB})-\mathcal{C}_r(Q,T,\rho_{AB})$, against the parameter \(p\) for the Werner state, $\rho_{AB}=p\frac{I}{4}+(1-p)\proj{\Psi}$, for the bases chosen in \emph{Example 2}. We see that the Bell-type inequality for relative entropy of coherence is violated (${\cal B}_{{\cal C}_{r}} >4$) for $0 \leq p \lesssim 0.07$.}
\label{fig:bell-cr-werner}
\end{center}
\end{figure}
Next, for the Werner state $\rho_{AB}=p\frac{I}{4}+(1-p)\proj{\Psi}$, for $Q,~R,~S,~T$ chosen in Example 2 above, we obtain
\begin{eqnarray*}
{\cal B}_{{\cal C}_{r}}=5-H({\frac{p}{2},1-\frac{p}{2}})-2H(\frac{p}{4},\frac{p}{4},\frac{p}{4},\frac{4-3p}{4}),
\end{eqnarray*}
where $H(\set{p_i})=\sum-p_i\log p_i$ is the Shannon entropy of the probability distribution of
$\set{p_i}$. The plot of the quantity ${\cal B}_{{\cal C}_{r}}$, in the above equation, against the white noise parameter \(p\) is shown in Fig. \ref{fig:bell-cr-werner}. We see that the Bell-type inequality for relative entropy of coherence is violated for $0 \leq p \lesssim 0.07$. 
Moreover, for $l_1$-norm of coherence,
\begin{eqnarray*}
{\cal B}_{{\cal C}_{l_1}}=10-8p.
\end{eqnarray*}
The value of the above inequality is larger than 8 if $p\in [0, \frac{1}{4})$.\\

Based on above observations, we ask the following question:
\emph{when Alice and Bob choose observables Q, R, S and T such that Bell inequality is violated and the maximal value $2\sqrt{2}$ is obtained, then do these observables violate Bell-type inequalities in coherence theory?} We find that this is not true as shown below.
As we know, when $\ket{\Psi}=\frac{1}{\sqrt{2}}(\ket{01}-\ket{10})$, and we take Q, R, S and T as the following \cite{Clauser1969}:
\begin{eqnarray}
Q=Z,&&~~R=X\\
S=(-Z-X)/\sqrt{2},&&~~T=(Z-X)/\sqrt{2}
\end{eqnarray}
then the Bell inequality attains the maximal value. The eigenstates of these observables are
\begin{eqnarray*}
&&Q: \set{\ket{0}, \ket{1}},~~R: \set{\ket{+}, \ket{-}}\\
&&S: \set{\ket{S_+}, \ket{S_-}},~~T: \set{\ket{T_+}, \ket{T_-}}
\end{eqnarray*}
where $\ket{S_+}=\frac{1}{\alpha}(\ket{+}+\ket{0})$ and $\ket{S_-}=\frac{1}{\beta}(\ket{+}-\ket{0})$ are the eigenstates of the observable $(-Z-X)/\sqrt{2}$, and $\ket{T_+}=\frac{1}{\alpha}(\ket{+}+\ket{1})$ and $\ket{T_-}=\frac{1}{\beta}(\ket{+}-\ket{1})$ are the eigenstates of the observable $(Z-X)/\sqrt{2}$, and
$\alpha=\sqrt{2+\sqrt{2}}$, $\beta=\sqrt{2-\sqrt{2}}$. After some simple calculations, we have
$\mathcal{C}_{l_1}(Q,S,\ket{\Psi})+1=\mathcal{C}_{l_1}(R,S,\ket{\Psi})+1
=\mathcal{C}_{l_1}(R,T,\ket{\Psi})+1=\mathcal{C}_{l_1}(Q,T,\ket{\Psi})+1=2+\sqrt{2}$, and
$\mathcal{C}_r(Q,S,\ket{\Psi})=\mathcal{C}_r(R,S,\ket{\Psi})=\mathcal{C}_r(R,T,\ket{\Psi})=\mathcal{C}_r(Q,T,\ket{\Psi})=H(\frac{1}{4\alpha^2}, \frac{1}{4\alpha^2}, \frac{1}{4\beta^2}, \frac{1}{4\beta^2})=1.6009$, where $H(\vec{p})= -\sum_i p_i \log p_i$ is the Shannon entropy.
Thus,
\begin{eqnarray*}
{\cal B}_{{\cal C}_{l_1}}=4+2\sqrt{2}<8,
\end{eqnarray*}
and
\begin{eqnarray*}
{\cal B}_{{\cal C}_{r}}=3.2018 < 4.
\end{eqnarray*}


\section{Bell-type inequality for Relative entropy of Coherence as an Entanglement Witness}
\label{sec:ent-witness}
Below we choose a set of observables Q, R, S and T, in whose basis, the Bell-type inequality for relative entropy of coherence is violated by all two-qubit entangled pure states, acting as an entanglement witness. We also find that, in the same basis, Bell-type inequality for $l_1$-norm of coherence is violated by a large number of two-qubit pure entangled states.

\begin{prop}
For any two-qubit entangled pure state,
$\ket{\psi}=\cos \frac{\theta}{2}\ket{00}+\sin\frac{\theta}{2} e^{i\phi}\ket{11}$, where
$\theta\in (0,\pi)$ and $\phi\in [0,2\pi]$, there exist observables Q, R in $\cH_A$ and S, T in $\cH_B$, such that
\begin{eqnarray*}
&&\mathcal{C}_r(Q,S,\rho_{AB})+\mathcal{C}_r(R,S,\rho_{AB})\nonumber\\
&+&\mathcal{C}_r(R,T,\rho_{AB})-\mathcal{C}_r(Q,T,\rho_{AB})> 4.
\end{eqnarray*}
\end{prop}

\begin{proof}
Consider the following basis:
\begin{eqnarray*}
&&Q: \set{\ket{0}, \ket{1}},~~R: \set{\ket{R_+}, \ket{R_-}}\\
&&S: \set{\ket{S_+}, \ket{S_-}},~~T: \set{\ket{0}, \ket{1}}
\end{eqnarray*}
where
\begin{eqnarray*}
\ket{R_+}=\frac{1}{\sqrt{2}}(\ket{0}+ie^{i\phi}\ket{1}),\\
\ket{R_-}=\frac{1}{\sqrt{2}}(\ket{0}-ie^{i\phi}\ket{1}),
\end{eqnarray*}
and
\begin{eqnarray*}
\ket{S_+}=\frac{1}{\sqrt{2}}(\ket{0}+\ket{1}),\\
\ket{S_-}=\frac{1}{\sqrt{2}}(\ket{0}-\ket{1}).
\end{eqnarray*}
For these observables, relative entropy of coherences are
$\mathcal{C}_r(R,S,\ket{\Psi})=2$, $\mathcal{C}_r(R,T,\ket{\Psi})=\mathcal{C}_r(Q,S,\ket{\Psi})=1+h\big(\cos^2\frac{\theta}{2},\sin^2\frac{\theta}{2}\big)$,
$\mathcal{C}_r(Q,T,\ket{\Psi})=h\big(\cos^2\frac{\theta}{2},\sin^2\frac{\theta}{2}\big)$,
where $h(p,1-p)=-p\log_2 p-(1-p)\log_2(1-p)$ is the Shannon entropy for the probability
distribution $\set{p,1-p}$. Therefore,
\begin{eqnarray*}
&&\mathcal{C}_r(Q,S,\ket{\Psi})+\mathcal{C}_r(R,S,\ket{\Psi})\\
&+&\mathcal{C}_r(R,T,\ket{\Psi})-\mathcal{C}_r(Q,T,\ket{\Psi})\\
&=&4+h\big(\cos^2\frac{\theta}{2},\sin^2\frac{\theta}{2}\big).
\end{eqnarray*}
Thus, for above set of observables $Q,~R,~S,~T$, we see that when $\ket{\psi}$ is entangled (i.e., $\theta \in (0,\pi)$), Bell-type inequality for relative entropy of coherence is violated, serving as an entanglement witness.
\end{proof}
However, the set of observables $Q,~R,~S,~T$ do not always violate the Bell-type inequality for $l_1$-norm of coherence, as shown below. For these observables, $l_1$-norm of coherences are
$\mathcal{C}_{l_1}(R,S,\ket{\Psi})+1=4$, $\mathcal{C}_{l_1}(R,T,\ket{\Psi})+1=\mathcal{C}_{l_1}(Q,S,\ket{\Psi})+1=2(|\cos\theta/2|+|\sin\theta/2|)^2$ and $\mathcal{C}_{l_1}(Q,T,\ket{\Psi})+1=(|\cos\theta/2|+|\sin\theta/2|)^2$.
Hence,
\begin{eqnarray*}
&&[\mathcal{C}_{l_1}(Q,S,\rho_{AB})+1]+[\mathcal{C}_{l_1}(R,S,\rho_{AB})+1]\\
&+&[\mathcal{C}_{l_1}(R,T,\rho_{AB})+1]-[\mathcal{C}_{l_1}(Q,T,\rho_{AB})+1]\\
&=&4+3(|\cos\theta/2|+|\sin\theta/2|)^2\\
&=&7+3\sin\theta.
\end{eqnarray*}
Thus, we see that when $\theta\in(\arcsin \frac{1}{3},\pi-\arcsin\frac{1}{3})$, the inequality is violated
\begin{eqnarray*}
&&[\mathcal{C}_{l_1}(Q,S,\rho_{AB})+1]+[\mathcal{C}_{l_1}(R,S,\rho_{AB})+1]\\
&+&[\mathcal{C}_{l_1}(R,T,\rho_{AB})+1]-[\mathcal{C}_{l_1}(Q,T,\rho_{AB})+1]>8,
\end{eqnarray*}
and when $\theta\in[0,\arcsin\frac{1}{3}]\cup[\pi-\arcsin\frac{1}{3},\pi]$, it does not exceed 8.

\section{Conclusion}
\label{sec:con}
Bell's theorem is regarded as one of the most profound discoveries of science. It is a `no-go theorem' that draws distinction between quantum mechanics and the world as described by classical mechanics.
It states that \emph{no physical theory of local hidden variables can ever reproduce all of the predictions of quantum mechanics}. Bell had constructed a mathematical inequality derived from locality and reality assumptions, that could be tested experimentally and showed it to be violated by certain
entangled quantum mechanical states. Violations of Bell inequalities form necessary
and sufficient criteria to detect entanglement in pure bipartite states. As entanglement arises from quantum coherence, in this work, we have investigated Bell-type inequalities for different valid measures of coherence.
First, we constructed Bell-type inequalities for $l_1$-norm of coherence and relative entropy of coherence, and found that these inequalities are violated by the maximally entangled two-qubit pure states. Furthermore, we showed that Bell-type inequality for relative entropy of coherence is violated by all two-qubit entangled pure states. Thus, we see that violation of the Bell-type inequality serve as an entanglement witness.
We conclude by stating some future directions that one could explore: we don't know whether these inequalities are violated by all two-qubit mixed states. Also, we don't know the maximal values of violation of the Bell-type inequalities that we have obtained. 

\begin{acknowledgments}
AK acknowledges the research fellowship of Department of Atomic Energy, Government of India. This project
is supported by National Natural Science Foundation of China (11171301, 11571307) and by the Doctoral Programs Foundation of the Ministry of Education of China (J20130061).
\end{acknowledgments}

\bibliographystyle{apsrev4-1}
 \bibliography{bell-coh-lit}

\end{document}